\documentstyle[prd,aps,preprint,tighten]{revtex}
\begin{document}
\draft 

\title{Octet baryon magnetic moments in the chiral quark model with
configuration mixing} \author{Johan Linde,\footnote{Electronic address:
jl@theophys.kth.se} Tommy Ohlsson,\footnote{Electronic address:
tommy@theophys.kth.se} and H{\aa}kan Snellman\footnote{Electronic
address: snell@theophys.kth.se}} \address{Theoretical Physics,
Department of Physics, Royal Institute of Technology, SE-100 44
Stockholm, Sweden} \date{Received 11 April 1997}

\maketitle

\begin{abstract}
The Coleman--Glashow sum-rule for magnetic moments is always
fulfilled in the chiral quark model, independently of SU(3)
symmetry breaking.  This is due to the structure of the wave
functions, coming from the non-relativistic quark model.
Experimentally, the Coleman--Glashow sum-rule is violated by about ten
standard deviations.  To overcome this problem, two models of wave
functions with configuration mixing are studied.  One of these models
violates the Coleman--Glashow sum-rule to the right degree and also
reproduces the octet baryon magnetic moments rather accurately.
\end{abstract}


\narrowtext

\section{Introduction}
\label{sec:intro}

The quark structure of baryons at low energies are probed by
parameters such as magnetic moments, axial-vector form factors and
decay rates of various kinds.  Any refinement of the non-relativistic
quark model (NQM) should improve on the experimental agreement of
these parameters, if the refinement is significant.  Much work has
been done to effectuate such refinements and improve the agreement
with the magnetic moments, the spin polarization of the nucleon, etc.
Among these refinements, the chiral quark model ({$\chi$}QM) suggested
by Manohar and Georgi \cite{mano84} has attracted some attention
recently \cite{eich92,chen95,lilf95,chen952,chen97,song97,webe97}.
Other models are one with quark-gluon configuration mixing by
Lipkin\cite{lipk90}, and one with quark-diquark configuration mixing
by Noda {\it et al.} \cite{noda94}.

One crucial test for quark model refinements is the Coleman--Glashow
sum-rule \cite{cole61} \[ \mu(p)-\mu(n) +\mu(\Sigma^-) -\mu(\Sigma^+)
+\mu(\Xi^0)-\mu(\Xi^-) =0 \] for the magnetic moments of the octet
baryons, that can be derived under very general assumptions on the
magnetic moment operator.  Experimentally, this sum-rule is violated
by ten standard deviations, the left hand side being equal to $(0.49
\pm 0.05)\,\mu_N$.

Franklin \cite{fran69,fran84} and Karl \cite{karl92} have shown that
the Coleman--Glashow sum-rule is valid beyond the NQM. Franklin noted
the validity of this sum-rule under the assumption of ``baryon
independence'' of a given quark moment contribution. Karl considered
the case of general quark spin polarizations and showed that the
sum-rule is valid assuming SU(3) symmetry for the wave functions of
the baryon octet states.

As we will show below, the Coleman--Glashow sum-rule turns out to hold
also in the $\chi$QM with arbitrary SU(3) symmetry breaking, as long
as the wave functions for baryons with $xxy$
quarks ($x,y = u,d,s$, $x \neq y$) have the same (mirror) symmetry.
This indicates a certain over-simplification in the description of the
baryons in this model and in several other models.

One possible way to remedy this is to allow the quark magnetic moments
to vary between the isomultiplets.  The alleged symmetry is then not
relevant.  This approach has the disadvantage of complicating the
quark model, by making the quarks vary with environment.  In fact, we
know that the mass spectrum can be well accounted for using the same
quark masses in all isomultiplets.  It is therefore desirable to
instead modify the wave functions, keeping the quark properties the
same throughout.

A natural modification of the mirror symmetry occurs when the quarks
are allowed to have an orbital angular momentum in the wave function.
The reason is that the mass of the $s$ quark breaks the symmetry.  An
example of such a model has been suggested by Casu and Sehgal
\cite{casu96}.  Using their formulas, the Coleman--Glashow sum-rule is
indeed violated and the left hand side is approximately given by $0.06
\, \langle L_z \rangle \, \mu_N$, where $\langle L_z \rangle$ is the
angular momentum.  To reach the experimental value of $0.49 \, \mu_N$,
this requires $\langle L_z \rangle$ to be about $8$, a value which is
unfortunately quite unrealistic.

Another model, which also breaks the Coleman--Glashow magnetic moment
sum-rule, is given by SU(3) breaking terms in a purely
phenomenological SU(3) parametrization \cite{bosc95,bosc96}.  This
model satisfies the experimental value for the left hand side.  On the
other hand, this model does not have any polarization of the vacuum,
and therefore the violation of the Gottfried sum-rule, giving $\bar{u}
- \bar{d} \simeq - 0.15$, cannot be explained.

Buck and Perez \cite{buck95} have discussed a model in which they add
a configuration term to the usual SU(6) spin function. This term
involves a total angular momentum of the quarks with $L=1$. Their
model violates the Coleman--Glashow sum-rule and gives $0.40\,\mu_{N}$
for the left hand side, but neither this model includes any vacuum
polarization.

In this paper we will therefore concentrate our further discussion to
the $\chi$QM and study two models of configuration mixing in the wave
functions of the octet baryons.

In the first model, this is done in the form of a gluon coupled to the
three quarks in a way suggested by Lipkin \cite{lipk90}.  The full
wave function, being a superposition of the one with zero gluons and
the one with one gluon, there is a natural room for varying the
relative importance of these two components for the different
isomultiplets.  This creates a breaking of the mirror symmetry that
generates the breaking of the Coleman--Glashow sum-rule.

In the other model, we use instead of a quark-gluon a quark-diquark
configuration mixing, that is allowed to vary between the
isomultiplets.

Both these models have been used originally without the Goldstone
bosons that play an essential role in the $\chi$QM. Their performance
is then not satisfactory in other respects, like the $\bar u$-$\bar d$
asymmetry.  In our paper, we use the mechanisms of these two models to
generate the configuration mixing needed to break the Coleman--Glashow
sum-rule.  This configuration mixing can be viewed as a correction to
the SU(6) quark model baryonic wave functions.  At the end of this
article, we will give an example, in the form of a toy model, how such
a configuration mixing could come about.

Our paper is organized as follows.  In Sec.~\ref{sec:CGSMM} we
first review the $\chi$QM, and then we show that the $\chi$QM with
arbitrary SU(3) symmetry breaking generates octet baryon magnetic
moments that satisfy the Coleman--Glashow sum-rule.  In
Sec.~\ref{sec:CQMCM} we then introduce two different models for
configuration mixing in the octet baryon wave functions, one with
quark-gluon mixing and one with quark-diquark mixing, and we show that
the Coleman--Glashow sum-rule can be violated in these models provided
that the mixings are allowed to vary between the isomultiplets.  At
the end of this section, we discuss a toy model for configuration
mixing.  Finally, in Sec.~\ref{sec:SC}, we present a summary of our
analyses and also the main conclusions.

\section{The Coleman--Glashow Sum-rule for Magnetic Moments}
\label{sec:CGSMM}

\subsection{The chiral quark model}

The Goldstone bosons (GBs) of the $\chi$QM are pseudoscalars and will
be denoted by the $0^{-}$ meson names $\pi,K,\eta,\eta'$, as is
usually done.  For convenience, we will closely follow the notation of
Ref.  \cite{chen95}.  The Lagrangian of interaction, ignoring the
space-time structure, is to lowest order
\begin{equation}
\label{eq:lag}
{\cal L_{\it I}} = g_8 \, \bar{{\bf q}} \, \Phi \, {\bf q},
\end{equation}
where $g_8$ is a coupling constant,
\begin{displaymath}
	{\bf q} = \left(\begin{array}{c} u \\ d \\ s
	\end{array}\right),
\end{displaymath}
 and
\begin{displaymath}
	\Phi = \left ( \begin{array}{ccc} c_{\pi^0}
	\frac{\pi^0}{\sqrt{2}} + c_{\eta} \frac{\eta}{\sqrt{6}} +
	c_{\eta'} \frac{\eta'}{\sqrt{3}} & c_{\pi^+} \pi^+ & c_{K^+}
	K^+\\ c_{\pi^-} \pi^- & - c_{\pi^0} \frac{\pi^0}{\sqrt{2}} +
	c_{\eta} \frac{\eta}{\sqrt{6}} + c_{\eta'}
	\frac{\eta'}{\sqrt{3}} & c_{K^0} K^0\\ c_{K^-} K^- &
	c_{\bar{K}^0} \bar{K}^0 & - c_{\eta} \frac{2 \eta}{\sqrt{6}} +
	c_{\eta'} \frac{\eta'}{\sqrt{3}}\\ \end{array} \right ),
\end{displaymath}
where all $c_i$ are parameters.

The effect of this coupling is that the emission of the GBs will
create quark-antiquark pairs from the vacuum with quantum numbers of
the pseudoscalar mesons.  Goldstone boson (GB) emission will therefore
in general flip the spin of the quarks.  The interaction of the GBs is
weak enough to be treated by perturbation theory.  This means that on
long enough time scales for the low energy parameters to develop we
have
\begin{mathletters}
\begin{eqnarray}
	u^{\uparrow} & \rightleftharpoons & (d^{\downarrow}+ \pi^+) +
	(s^{\downarrow} +K^+) + (u^{\downarrow} + \pi^0, \eta, \eta'),
	\\ d^{\uparrow}& \rightleftharpoons & (u^{\downarrow}+ \pi^-)
	+ (s^{\downarrow} +K^0 ) + (d^{\downarrow} + \pi^0 ,\eta
	,\eta'), \\ s^{\uparrow} & \rightleftharpoons &
	(u^{\downarrow}+ K^-) + (d^{\downarrow} +\bar{K}^0) +
	(s^{\downarrow} + \eta ,\eta').
\end{eqnarray}
\end{mathletters}

The matrix $\Phi$ in the Lagrangian~(\ref{eq:lag}) is the most general
parametrization of the pseudoscalar GB matrix in the $\chi$QM. In a
realistic model, one should of course not use all these parameters.
The reason for introducing this large set of parameters is to make the
following discussion general.  The parameter $c_{\eta'}$ describes
U(3) symmetry breaking and the other parameters describe SU(3)
symmetry breaking.

Cheng and Li have used the SU(3) symmetric model with a broken U(3)
symmetry \cite{chen95} and showed that it can successfully be used to
calculate the quark spin polarizations in the nucleon.  In a later
paper\cite{chen97}, they have extended this model by introducing SU(3)
symmetry breaking in the Lagrangian via two parameters $c_{K}=\alpha$
and $c_{\eta}=\beta$.  Song {\it et al.} \cite{song97} and Weber {\it
et al.} \cite{webe97} have also studied models with SU(3) symmetry
breaking, similar to the one discussed by Cheng and Li.  All these
extended models have lead to significantly better results for several
physical quantities.

\subsection{The Coleman--Glashow sum-rule}

There is, however, one important set of data which the $\chi$QM cannot
successfully predict regardless how many symmetry breaking parameters
one introduces in the Lagrangian~(\ref{eq:lag}): the octet baryon
magnetic moments.  This is the case at least as long as one uses SU(6)
symmetric wave functions for the octet baryons.  This is most easily
illustrated by the function
\begin{equation}
	\Sigma_{\mu}\equiv\mu(p) - \mu(n) + \mu(\Sigma^-) -
	\mu(\Sigma^+) + \mu(\Xi^0) - \mu(\Xi^-).
\label{summaregel}
\end{equation}
Experimentally, $\Sigma_{\mu}=(0.49\pm0.05) \, \mu_N$, but, as we will
show, in the $\chi$QM $\Sigma_{\mu}=0$ (the Coleman--Glashow
sum-rule).

Writing out the explicit valence quark content of the baryons in
Eq.~(\ref{summaregel}) we have
\begin{equation}
	\Sigma_{\mu}=\mu{\bf (}B(uud){\bf )}-\mu{\bf (}B(ddu){\bf )}+
	\mu{\bf (}B(dds){\bf )}-\mu{\bf (}B(uus){\bf )}+\mu{\bf
	(}B(ssu){\bf )}-\mu{\bf (}B(ssd){\bf )}.
\end{equation}
To obtain $\Sigma_{\mu}=0$ we need a
mirror symmetry, such that the contribution to the magnetic moment
generated by GB emission from the two $u$ quarks in
$B(uud)$ cancels the corresponding contribution generated
by GB emission from the two $u$ quarks in $B(uus)$, the
contribution generated by the $d$ quark in $B(uud)$ cancels the one
generated by the
$d$ quark in $B(ssd)$, etc., provided that the quark magnetic moments
are constant.  This is trivially true in the NQM. As mentioned in the
Introduction, there is a large class of models beyond the NQM, where the
Coleman--Glashow sum-rule is fulfilled \cite{fran69,fran84,karl92}.  
We will now make a schematic calculation to show that the above condition is
fulfilled in the $\chi$QM with arbitrary SU(3) symmetry breaking.

First, we introduce a function $\hat{B}$ to describe the spin
structure of a baryon $B$
\begin{equation}
	\hat{B}=n_{x^{\uparrow}}\hat{x}^{\uparrow}+n_{x^{\downarrow}}
	\hat{x}^{\downarrow}+
	n_{y^{\uparrow}}\hat{y}^{\uparrow}+n_{y^{\downarrow}}
	\hat{y}^{\downarrow}+
	n_{z^{\uparrow}}\hat{z}^{\uparrow}+n_{z^{\downarrow}}
	\hat{z}^{\downarrow}.
\label{b-hatt}
\end{equation}
The coefficient $n_{q^{\uparrow\downarrow}}$ of each symbol
$\hat{q}^{\uparrow\downarrow}$ should be interpreted as the number of
$q^{\uparrow\downarrow}$ quarks.  See Appendix~\ref{app:survey} for a
complete discussion of the function $\hat{B}$.  Then, $\Delta
q^{B}=n_{q^{\uparrow}}(B)-n_{q^{\downarrow}}(B)$ is the $q$ quark spin
polarization in the baryon $B$.  Normally, there is also a
contribution from the antiquarks to the spin polarization, but in the
$\chi$QM this is zero.  The baryon magnetic moments can be
parametrized as
\begin{equation}
	\mu(B)=\Delta u^{B}\mu_{u}+\Delta d^{B}\mu_{d}+\Delta
	s^{B}\mu_{s}, \label{eq:bar_mom}
\end{equation}
where $\mu_{q}$ is the quark magnetic moment of the $q$ quark.  Here
the quark spin polarization, $\Delta q^{B}$, may vary from baryon to
baryon, but the quark magnetic moment, $\mu_q$, is the same for all
baryons.

The starting point in the $\chi$QM is the spin structure in the NQM.
The NQM spin structure of an octet baryon $B(xxy)$ is
\begin{equation}
	\hat{B}(xxy)
	=\frac{5}{3}\hat{x}^{\uparrow}+\frac{1}{3}\hat{x}^{\downarrow}
	+\frac{1}{3}\hat{y}^{\uparrow}+\frac{2}{3}\hat{y}^{\downarrow},
	\label{baryonmoment}
\end{equation}
so the spin polarizations are $\Delta x^B = \frac{4}{3}$, $\Delta y^B
= - \frac{1}{3}$, and $\Delta z^B = 0$, where $z$ is the non-valence
quark.  Using this it is easy to see that the Coleman--Glashow
sum-rule is fulfilled in the NQM. With help of
Eq.~(\ref{baryonmoment}) we can express the spin structure after one
iteration in the $\chi$QM by
\begin{eqnarray}
	\hat{B}(xxy)=&& P_x
	\left(\frac{5}{3}\hat{x}^{\uparrow}+\frac{1}{3}
	\hat{x}^{\downarrow}\right) + P_y
	\left(\frac{1}{3}\hat{y}^{\uparrow} +
	\frac{2}{3}\hat{y}^{\downarrow}\right) \nonumber \\ && {} +
	\frac{5}{3} \vert \psi(x^{\uparrow}) \vert^2 + \frac{1}{3}
	\vert \psi(x^{\downarrow}) \vert^2 + \frac{1}{3} \vert
	\psi(y^{\uparrow}) \vert^2 +\frac{2}{3} \vert
	\psi(y^{\downarrow}) \vert^2, \label{eq:8}
\end{eqnarray}
where $P_{q}$ is the probability of no GB emission from the $q$ quark
and $\vert \psi(q^{\uparrow \downarrow}) \vert^2$ are the
probabilities of GB emission from the $q^{\uparrow \downarrow}$
quarks.  The functions $P_q$ and $\vert \psi(q^{\uparrow \downarrow})
\vert^2$ are discussed in detail in Appendix~\ref{app:survey}.

For example, the probability function $\vert \psi(x^\uparrow) \vert^2$
is of the form
\begin{equation}
	\vert \psi(x^{\uparrow}) \vert^2 = b_{x^{\downarrow}}
	\hat{x}^{\downarrow} + b_{y^{\downarrow}} \hat{y}^{\downarrow}
	+ b_{z^{\downarrow}} \hat{z}^{\downarrow},\label{eq:9}
\end{equation}
where $b_{x^{\downarrow}}$, $b_{y^{\downarrow}}$, and
$b_{z^{\downarrow }}$ are some constants depending on the choice of
the parameters $c_{i}$ in the Lagrangian.  We have here omitted the
quark-antiquark pair created by the GB as it will not contribute to
the spin polarizations.

It is now easy to see that the sum-rule is fulfilled. 
For example, the two valence $u$ quarks in $B(uud)$ give a
contribution to the spin structure after GB emission, which is 
\begin{equation}
	P_u \left(\frac{5}{3}\hat{u}^{\uparrow}+\frac{1}{3}
	\hat{u}^{\downarrow}\right) + \frac{5}{3} \vert
	\psi(u^{\uparrow}) \vert^2 + \frac{1}{3} \vert
	\psi(u^{\downarrow}) \vert^2. \label{eq:10}
\end{equation}
This is canceled by an identical contribution from the $u$ quarks in
$B(uus)$.  Similarly, the contribution from the $d$ quark in $B(uud)$
\begin{equation}
	P_d \left(\frac{1}{3}\hat{d}^\uparrow + \frac{2}{3}
	\hat{d}^\downarrow\right) + \frac{1}{3}
	\vert\psi(d^\uparrow)\vert^2 + \frac{2}{3}
	\vert\psi(d^\downarrow)\vert^2
\end{equation}
will cancel the contribution from the $d$ quark in $B(ssd)$, etc.  This
shows that the Coleman--Glashow sum-rule $\Sigma_\mu = 0$ is satisfied
in the $\chi$QM with arbitrary symmetry breaking in the
Lagrangian~(\ref{eq:lag}).  One can also easily show that the
Coleman--Glashow sum-rule is fulfilled for arbitrary number of
iterations of GB emission in the $\chi$QM.

Note that expression~(\ref{eq:10}) contains a part of the spin polarization
of all three quarks, $u$, $d$, and $s$, as can be seen from
Eq.~(\ref{eq:9}). Similarly, the original $d$ quark in the proton
contributes by GB emission to the spin polarization of all three quarks. The 
contribution to the spin polarization of the $u$ quark generated by
the original
$d$ quark in the proton is in general different from the one generated
by the $s$ quark
in $\Sigma^+$, due to the symmetry breaking in the Lagrangian. 
This means that in general $\Delta u^p \neq \Delta
u^{\Sigma^+}$. Therefore the sum-rule is fulfilled only
because of the mirror symmetry in the NQM wave functions used as input
in Eq.~(\ref{eq:8}). The sum-rule is not a result of
baryon independent quark spin polarizations, but a result of the fact
that the total contribution from all six baryons to a given flavor
cancels. Thus, we have the relation
\begin{equation}
\Delta q^p - \Delta q^n + \Delta q^{\Sigma^-} - \Delta q^{\Sigma^+} +
\Delta q^{\Xi^0} - \Delta q^{\Xi^-} = 0,\quad q=u,d,s,
\end{equation}
rather than simple relations as {\it e.g.} $\Delta u^p = \Delta
u^{\Sigma^+}$. This can also be seen from the explicit expressions in
Appendix~\ref{app:spin_pol} (when $\theta_N = \theta_\Sigma = \theta_\Xi
= 0$).

\section{The Chiral Quark Model with Configuration Mixing}
\label{sec:CQMCM}

As we have shown above, the Coleman--Glashow sum-rule is satisfied in
the $\chi$QM. There are in principle two ways of overcoming this
problem as discussed before, one is to let the quark magnetic moments
vary between the isomultiplets of the octet baryons, and the other one
is to introduce symmetry breaking in the wave functions.  For reasons
discussed in Sec.~\ref{sec:intro}, we will here adopt the second
alternative.  One way of doing this is to add configuration mixing
terms in the wave functions.

In our models, the wave functions will have the general structure
\begin{equation}
\vert B^\uparrow \rangle \equiv \vert B; {\textstyle S = \frac{1}{2},
S_z = +\frac{1}{2} } \rangle = \cos \theta_B \vert B_{1}^{\uparrow}
\rangle + \sin \theta_B \vert {B'_1}^\uparrow \rangle,
\label{blandad_vag_funk}
\end{equation}
where $\vert B_{1}^{\uparrow} \rangle$ is the usual SU(6) wave
function and $\vert {B'_1}^\uparrow \rangle$ is the configuration
mixing term.  The angle $\theta_B$ is a measure of the amount of
mixing.  We will let the angles of configuration mixing be the same
within each baryon isomultiplet, but let them vary between different
isomultiplets.  We also assume, for simplicity, that the mixing angle
for $\Lambda$ is equal to the one for $\Sigma$.  Thus we have three
mixing angles $\theta_{N}$, $\theta_{\Sigma}$, and $\theta_{\Xi}$.

\subsection{Wave functions with quark-gluon mixing}

First, we will discuss a simple model with a wave function with an
additional term where a color-octet baryon state is coupled to a spin
one color-octet gluon state.  We call this model the chiral quark
model with quark-gluon mixing ($\chi$QM$g$).

The wave function for the octet baryons in this model is a mixture of
two different wave functions \cite{lipk90}
\begin{equation}
\vert B^\uparrow \rangle = \cos \theta_B \vert B_1^\uparrow \rangle +
\sin \theta_B \vert (B_8 G)^\uparrow \rangle.
\end{equation}
Thus, in this case we have set $|{B'_{1}}^{\uparrow}\rangle =
|(B_{8}G)^{\uparrow}\rangle$ in Eq.~(\ref{blandad_vag_funk}).

The octet baryon color-singlet wave function for the $xxy$ baryons is
given by
\begin{equation}
\vert B_1^\uparrow (xxy) \rangle = \frac{1}{\sqrt{6}} \left( 2 \vert
x^\uparrow x^\uparrow y^\downarrow \rangle - \vert x^\uparrow
x^\downarrow y^\uparrow \rangle - \vert x^\downarrow x^\uparrow
y^\uparrow \rangle \right)
\label{eq:b1xxy}
\end{equation}
and for the $\Lambda$ baryon by
\begin{equation}
\vert \Lambda_1^\uparrow (uds) \rangle = \frac{1}{\sqrt{2}} \left(
\vert u^\uparrow d^\downarrow s^\uparrow \rangle - \vert u^\downarrow
d^\uparrow s^\uparrow \rangle \right).
\label{eq:l1uds}
\end{equation}
We have suppressed color and permutations in flavor in the above wave
functions.  We will do so also in the following, as this will not
affect the spin structures.

The gluonic octet baryon color-singlet wave function is a coupling of
an octet baryon color-octet wave function, $\vert B_8 \rangle$, and a
spin-one color-octet gluon wave function, $\vert G \rangle$, to make a
color-singlet state with total angular momentum $J = \frac{1}{2}$
\begin{eqnarray}
\vert (B_8 G)^\uparrow \rangle &=& -\frac{1}{\sqrt{3}} \vert B_8;
{\textstyle S = \frac{1}{2}, S_z = +\frac{1}{2} } \rangle \otimes
\vert G; {\textstyle S = 1, S_z = 0 } \rangle \nonumber \\ &+&
\sqrt{\frac{2}{3}} \vert B_8; {\textstyle S = \frac{1}{2}, S_z =
-\frac{1}{2} } \rangle \otimes \vert G; {\textstyle S = 1, S_z = +1 }
\rangle.
\label{eq:b8g}
\end{eqnarray}
Here
\begin{equation}
\vert B_8 (xxy) ; {\textstyle S = \frac{1}{2}, S_z = +\frac{1}{2} }
\rangle = \frac{1}{\sqrt{3}} \left( \vert x^\uparrow x^\uparrow
y^\downarrow \rangle + \vert x^\uparrow x^\downarrow y^\uparrow
\rangle + \vert x^\downarrow x^\uparrow y^\uparrow \rangle \right)
\label{eq:b8xxy}
\end{equation}
for the $xxy$ baryons and
\begin{equation}
\vert \Lambda_8 (uds) ; {\textstyle S = \frac{1}{2}, S_z = +
\frac{1}{2} } \rangle = \frac{1}{\sqrt{2}} \left( \vert u^\uparrow
d^\downarrow s^\uparrow \rangle - \vert u^\downarrow d^\uparrow
s^\uparrow \rangle \right)
\label{eq:l8uds}
\end{equation}
for the $\Lambda$ baryon.

\subsection{Wave functions with quark-diquark mixing}

An alternative to the quark-gluon mixing as a source of configuration
mixing is given by a model with quark-diquark mixing.  We call this
model the chiral quark model with quark-diquark mixing ($\chi$QM$d$).

The diquark model is a modification of the usual quark model by
considering two quarks glued together to form a diquark.  There are
SU(3) sextet axial-vector diquarks and SU(3) triplet scalar diquarks.
We will only consider scalar diquarks.  The symbol $(q_1 q_2)_d$ will
denote a scalar diquark consisting of the quarks $q_1$ and $q_2$.

It has been suggested in Ref.~\cite{lich69}, that a quark-diquark
model can be used to calculate strong and electromagnetic properties
of baryons.  In such a model, the diquark, although formed as a bound
state of two quarks, is regarded as essentially elementary in its
interaction with a quark to form a baryon.

In this model, the wave function for the octet baryons is a mixture of
the usual SU(6) wave function and a quark-diquark wave function
\cite{noda94}
\begin{equation}
\vert B^\uparrow \rangle = \cos \theta_B \vert B_1^\uparrow \rangle +
\sin \theta_B \vert B_d^\uparrow \rangle.
\end{equation}
Thus, in this case we use
$|{B'_{1}}^{\uparrow}\rangle=|B_{d}^{\uparrow}\rangle$ in
Eq.~(\ref{blandad_vag_funk}).

The octet baryon color-singlet wave function for the $xxy$ baryons is
again given by Eq.~(\ref{eq:b1xxy}) and for the $\Lambda$ baryon by
Eq.~(\ref{eq:l1uds}).  The quark-diquark octet baryon wave function is
\begin{equation}
\vert B_d^\uparrow (xxy) \rangle \equiv \vert x^\uparrow \rangle
\otimes \vert (x y)_d \rangle = \vert x^\uparrow (x y)_d \rangle
\label{eq:bd}
\end{equation}
for the $xxy$ baryons and
\begin{equation}
\vert \Lambda_d^\uparrow (uds) \rangle = \frac{1}{\sqrt{6}} \left(
\vert u^\uparrow (d s)_d \rangle - \vert d^\uparrow (u s)_d \rangle -
2 \vert s^\uparrow (u d)_d \rangle \right)
\end{equation}
for the $\Lambda$ baryon \cite{jako93}.

\subsection{Discussion of parameters}

In our further calculations, we will use the following parameters in
the Lagrangian~(\ref{eq:lag}): $c_{\pi^0} = c_{\pi^+} = c_{\pi^-} =
1$, $c_{K^+} = c_{K^-} = c_{K^0} = c_{\bar{K}^0} = \alpha$, $c_\eta =
\beta$, and $c_{\eta'} = \zeta$.  In some of our calculations we will
use an SU(3) symmetric Lagrangian with $\alpha=\beta=1$. See
Appendix~\ref{app:survey} for a detailed discussion of the Lagrangian.

The parameter $a$, describing the probability of GB emission, and the
parameter $\zeta$ can be estimated from the $\bar{u}$-$\bar{d}$
asymmetry.  The New Muon Collaboration (NMC) experiment has measured
the isospin asymmetry difference of the quark sea in the proton to be
\cite{amau91,arne94}
\begin{equation}
\bar{u} - \bar{d} \simeq - 0.15.
\label{eq:ud1}
\end{equation}
In the $\chi$QM this difference is given by
\begin{equation}
\bar{u} - \bar{d} = a \left( \frac{2 \zeta + \beta}{3} - 1 \right).
\label{eq:ud2}
\end{equation}
The expressions for the antiquark numbers $\bar{u}$ and $\bar{d}$ are
given in Appendix~\ref{app:survey}.  Combining Eqs.~(\ref{eq:ud1}) and
(\ref{eq:ud2}) we obtain
\begin{equation}
a \simeq \frac{0.44}{3 - 2 \zeta - \beta}.
\label{eq:a}
\end{equation}
Similarly to Eq.~(\ref{eq:ud2}) we have for the antiquark density
ratio
\begin{equation}
\bar{u} / \bar{d} = \frac{21 + 2(2\zeta+\beta)+(2\zeta+\beta)^{2}}{33
- 2(2\zeta+\beta)+(2\zeta+\beta)^{2}}
\label{eq:u/d}
\end{equation}
with the experimental value $\bar{u} / \bar{d} = 0.51 \pm 0.09$
\cite{bald94}.

If we set $\beta = 1$, then Eq.~(\ref{eq:u/d}) reduces to
\begin{equation}
{\bar{u} / \bar{d}} \, \vert_{\beta = 1} = \frac{6 + 2 \zeta +
\zeta^2} {8 + \zeta^2}.
\end{equation}
{}From this we obtain $- 4.3 < \zeta < - 0.7$.  Following Cheng and Li
\cite{chen95}, we choose the value $\zeta = - 1.2$.  The value of $a$
is now given by Eq.~(\ref{eq:a}) to be $a \approx 0.10$, which is in
good agreement with Ref.~\cite{eich92}.  However, when $\beta$ is a
free parameter in the calculations, we have to use the relation $2
\zeta + \beta \simeq -1.4$, which comes from Eq.~(\ref{eq:a}), in
order to keep $a \approx 0.10$.  We therefore make the assumption that
\begin{equation}
	\zeta = - 0.7 - \frac{\beta}{2}.
\end{equation}
This also fixes the value of $\bar{u} / \bar{d}$ to $0.53$.

In what follows, we consider the case where the magnetic moments of
the quarks satisfy the relations
\begin{equation}
\label{eq:muu}
\mu_u = -2 \mu_d
\end{equation}
and
\begin{equation}
\label{eq:mus}
\mu_s = \frac{2}{3} \mu_d.
\end{equation}

\subsection{Numerical results}

As we have seen, when $\theta_N = \theta_\Sigma = \theta_\Xi = 0$,
$\Sigma_\mu = 0$ for every choice of the parameters $c_i$.  Also when
the mixing angles are the same, but not equal to zero, $\Sigma_\mu =
0$.  However, when at least one of the mixing angles $\theta_B$ is
different from the others, the value of $\Sigma_\mu$ will be non-zero.

In the models, that we will discuss, all three mixing angles
$\theta_N$, $\theta_\Sigma$, and $\theta_\Xi$ will be free parameters.
The magnetic moment of the $d$ quark, $\mu_d$, will also be a free
parameter and the other quark magnetic moments are then given by the
relations~(\ref{eq:muu}) and (\ref{eq:mus}).  In order to calculate
the magnetic moments of the octet baryons we will also need the quark
spin polarizations, which are obtained from the quark spin structures.
A detailed derivation of the spin polarizations starting from the
Lagrangian~(\ref{eq:lag}) can be found in Appendices~\ref{app:survey}
and \ref{app:spin_pol}.  The baryon magnetic moments are given by
Eq.~(\ref{eq:bar_mom}).

We fit the experimental data for the octet baryon magnetic moments and
the weak axial-vector form factor $g_A$.  Since the magnetic moments
depend on the products of quark magnetic moments and quark spin
polarizations, the use of $g_A$ serves as a normalization of the
parameters.  The parameter values obtained from the different fits can
be found in Table~\ref{tab:fit_data}.

Let us first say a few words about the NQM with configuration mixing,
{\it i.e.\/} no GB emission $(a=0)$.

In the case with quark-gluon mixing, we will get the NQM$g$, an
extension of the model for the proton suggested by Lipkin
\cite{lipk90}.  The NQM$g$ gives $\Sigma_\mu \approx 0.17 \, \mu_N$.
However, the NQM$g$ does not give rise to any $\bar{u}$-$\bar{d}$
asymmetry, because of lack of vacuum polarization.

In the case with quark-diquark mixing, we will get the NQM$d$, an
extension of the model for the proton considered by Noda {\it et al.}
\cite{noda94}.  This model gives a much better value on $\Sigma_\mu$,
than the NQM$g$.  The value obtained is $\Sigma_\mu \approx 0.36 \,
\mu_N$, which is still not within the experimental errors.  As in the
NQM$g$, there is no $\bar{u}$-$\bar{d}$ asymmetry in the NQM$d$.  The
mixings also become unrealistically large, for example
$\sin^{2}\theta_{\Sigma} \approx 0.65$.

We now continue with the $\chi$QM. We will discuss two cases, one
where we put $\alpha=\beta=1$, and one where we let $\alpha$ and
$\beta$ vary independently.  Thus in the first case, we have the
original SU(3) symmetric Lagrangian and we can study the effect of the
mixing angles alone.  The second case makes it possible to see how the
combination of symmetry breaking and wave function mixing improves the
results.

In the first calculation with $\alpha=\beta=1$, we obtain the mixings
$\sin^{2}\theta_{N} \approx 0.00$, $\sin^{2}\theta_{\Sigma} \approx
0.05$, and $\sin^{2}\theta_{\Xi} \approx 0.11$ in the quark-gluon
model, and $\sin^{2}\theta_N \approx 0.00$, $\sin^{2}\theta_\Sigma
\approx 0.25$, and $\sin^{2}\theta_\Xi \approx 0.33$ in the
quark-diquark model.  In Table~\ref{tab:octet} the values of the octet
baryon magnetic moments, $g_A$, and $\Sigma_\mu$ are presented
together with the experimental values.  The over all fit is obviously
better than in the case without mixing, as we have more parameters,
but the important result is that we are now able to obtain non-zero
values of the function $\Sigma_{\mu}$.  In the quark-gluon model we
obtain $\Sigma_\mu \approx 0.28 \, \mu_N$, which still differs from
the experimental value, but in the quark-diquark model we obtain
$\Sigma_{\mu}\approx 0.55 \, \mu_N$, which is very close to
experiment.  Note, in Table~\ref{tab:data}, that the total spin
polarization $\Delta\Sigma$ is the same in the $\chi$QM$g$ and
$\chi$QM$d$ as in the $\chi$QM, simply because the mixing angle
$\theta_N$ is zero in these models.

In the second calculation we also let $\alpha$ and $\beta$ be free
parameters.  In this fit we have to use $\zeta = -0.7 - \beta/2$, in
order to keep $a \approx 0.10$.  The values of the magnetic moments
are over all improved compared to the above case with
$\alpha=\beta=1$, especially $\chi^2$ decreases with a factor of about
10 in the $\chi$QM$d$, see Table~\ref{tab:octet}.  The value of
$\Sigma_\mu$ in the $\chi$QM$g$ is about the same as in the case with
$\alpha=\beta=1$, but in the $\chi$QM$d$ we obtain $\Sigma_\mu \approx
0.52 \, \mu_N$, which lies within the experimental errors.  The
symmetry breaking in the Lagrangian becomes relatively large, $\alpha
\approx 0.70$ and $\beta \approx 0.73$ in the quark-gluon model and
$\alpha \approx 0.69$ and $\beta \approx 0.55$ in the quark-diquark
model.  The values obtained for $\alpha$, which is a suppression
factor for kaon GB emission, are reasonable as it can be argued that
$\alpha$ is proportional to $m/m_{s}=2/3$
\cite{chen97,song97,webe97}. On the other hand, the mixing angles are
not changed very much compared to the fits with $\alpha=\beta=1$,
except for $\theta_{N}$, which gets a non-zero value in the
quark-diquark model.  The mixings obtained are $\sin^2 \theta_N
\approx 0.00$, $\sin^2 \theta_\Sigma \approx 0.25$, and $\sin^2
\theta_\Xi \approx 0.33$ in the $\chi$QM$g$ and $\sin^2 \theta_N
\approx 0.11$, $\sin^2 \theta_\Sigma \approx 0.34$, and $\sin^2
\theta_\Xi \approx 0.41$ in the $\chi$QM$d$.

By letting $\alpha$ and $\beta$ vary, the major improvement we obtain
is very good values for the weak axial-vector form factor, $g_A
\approx 1.26$ in the $\chi$QM$g$ and $g_A \approx 1.24$ in the
$\chi$QM$d$.  On the other hand, the total spin polarization
$\Delta\Sigma$ becomes somewhat large (see Table~\ref{tab:data}).

How does the choice of parametrization of the Lagrangian influence
the results? We have chosen to introduce the SU(3) symmetry breaking
parameters $\alpha$ and $\beta$ in the same spirit as has been done by
other authors\cite{chen97}.  There are of course other options. For
example, it is possible that the probability of $d\to K^{0}+s$ is
different from that for $s\to \bar{K}^{0}+d$ due to the different
phase space. Taking this into account would require a set of new
parameters. Although this would give small corrections to the results,
it would not change the main conclusions. As has been pointed out,
there is no way to break the Coleman--Glashow sum-rule in the
$\chi$QM$d$ by introducing more symmetry breaking parameters in the
Lagrangian.

To investigate how a different set of parameters would influence the
results, we have considered the case where the substitution
$\varphi_{sq} \to \epsilon \varphi_{sq}$, $q = u,d$ has been carried
out in the last row in the matrix (\ref{Phihatt}).  The parameter
$\epsilon$ accounts for the difference in probability of an $s$ quark
emitting a GB and a $u$ or $d$ quark emitting a GB, as discussed
above. We have made a fit including $\epsilon$ in the model
$\chi$QM$d$, when $\alpha$ and $\beta$ are considered as free
parameters.  For $\epsilon$ we obtained the value 1.27. This results
in minor changes of the parameters $\alpha$ and $\beta$. The mixing
angles are $\sin^2 \theta_N \approx 0.10$, $\sin^2 \theta_\Sigma
\approx 0.32$, and $\sin^2 \theta_\Xi \approx 0.40$, which are almost
identical to the fit with $\epsilon = 1$ (see the last column in
Table~\ref{tab:fit_data}). This shows that the exact choice of
parametrization in the Lagrangian does not affect the main
conclusions, and verifies that the introduction of further SU(3)
breaking parameters in the $\chi$QM Lagrangian cannot reduce the size
of configuration mixing needed to break the Coleman--Glashow sum-rule.

\subsection{A Simple Mechanism for Configuration Mixing}
\label{sub:mechanism}

We will here describe a simple mechanism in the form of a toy model
for configuration mixing in the wave functions for the octet baryons.

In this simple toy model, we assume that we have a two level system of
mass states for the octet baryons, such that these are mixings of
(1) the usual three quark mass states, where the $u$ and $d$ quarks
have mass $m$ and the $s$ quark has mass $m_s$, and (2)
quark-diquark mass states.  The mass of the $(ud)_d$ diquark is $M$
and the mass of $(us)_d$ and $(ds)_d$ diquarks is $M_s$.  The diquarks
are only singlets.
When these states mix we obtain the two physical mass states, the
ground state and the first excited state.  The first excited state is
simply assumed to be the mass state next in order to the ground state
with the same quantum numbers as the ground state.  Thus, we interpret
the excitation to be a quark-diquark excitation rather than a radial
excitation.

The wave functions $\Psi_-$ and $\Psi_+$, corresponding to the
physical mass states, can be expressed in the wave functions $\Psi_0$
and $\Psi_1$, corresponding to the unphysical mass states, as
\begin{equation}
\left\{ \begin{array}{l} \Psi_- = \Psi_0 \cos \vartheta + \Psi_1 \sin
\vartheta\\ \Psi_+ = - \Psi_0 \sin \vartheta + \Psi_1 \cos \vartheta
\end{array} \right.
\label{eq:psi}
\end{equation}
where $\vartheta$ is the configuration mixing angle.  The wave
function $\Psi_-$ should be compared to Eq.~(\ref{blandad_vag_funk}).

We then introduce the Hamiltonian
\begin{equation}
\hat{H} = \left( \begin{array}{cc} m_{0} & H \\ H & m_{1} \end{array}
\right),
\end{equation}
where $m_0$ is the lower unphysical mass state and $m_1$ is the higher
unphysical mass state.  The parameter $H$ corresponds to the
transition probability between the unphysical mass states $m_0$ and
$m_1$, and it is assumed to be the same for $N$, $\Lambda$, $\Sigma$,
and $\Xi$.

For the lower unphysical mass states we use a simple mass formula with
a hyperfine coupling term \cite{grif87}
\begin{equation}
	m_{0}{\bf (}B(q_1 q_2 q_3){\bf )}=m_{q_1}+m_{q_2}+m_{q_3}+
	h\left(\frac{{\bf s}_{q_{1}}\cdot{\bf
	s}_{q_{2}}}{m_{q_{1}}m_{q_{2}}} +\frac{{\bf
	s}_{q_{1}}\cdot{\bf s}_{q_{3}}}{m_{q_{1}}m_{q_{3}}}+
	\frac{{\bf s}_{q_{2}}\cdot{\bf
	s}_{q_{3}}}{m_{q_{2}}m_{q_{3}}}\right),
\end{equation}
where ${\bf s}_{q_i}$ is the spin of the quark $q_i$.  The parameter
$h$ is the QCD hyperfine coupling parameter.  Since the diquarks are
scalars, there is no hyperfine coupling in the higher unphysical mass
states.  The mass formulas for the higher unphysical mass states are
$m_1(N) = m + M$, $m_1(\Lambda) = (m+M_s)/3 + 2(m_s+M)/3$,
$m_1(\Sigma) = m + M_s$, and $m_1(\Xi) = m_s + M_s$.

Solving the eigenvalue problem for the Schr{\"o}dinger equation $\hat{H}
\Psi = E \Psi$, where 
$$
\Psi = \left( \begin{array}{c} \Psi_0 \\ \Psi_1
\end{array} \right),
$$
we get the eigenvalues
\begin{equation}
E_{\pm} = \frac{m_0 + m_1}{2} \pm \frac{1}{2} \sqrt{(m_0-m_1)^2 + 4
H^2},
\end{equation}
which should correspond to the physical mass states.

The quantity $\sin^2 \vartheta$ measures the part of the total mass
state which is of quark-diquark origin and is given by
\begin{equation}
\sin^2 \vartheta = \frac{2 x^2}{1 + 4 x^2 + \sqrt{1+ 4 x^2}},
\end{equation}
where $x = H/(m_1 - m_0)$.

Choosing the illustrative values $m = 400 \;{\rm MeV}$, $m_s = 590
\;{\rm MeV}$, $M =920 \;{\rm MeV}$, $M_s = 1000 \;{\rm MeV}$, $h/(4
m^2) = 60 \;{\rm MeV}$, and $H = 190 \;{\rm MeV}$, we obtain an octet
baryon mass spectrum, which is in good agreement with the measured
spectrum.  For the mixings we get $\sin^2 \vartheta_N \approx 0.19$,
$\sin^2 \vartheta_\Lambda \approx 0.22$, $\sin^2 \vartheta_\Sigma
\approx 0.36$, and $\sin^2 \vartheta_\Xi \approx 0.36$.  The mixing
for $\Sigma$ and $\Xi$ is the same, since $m_1(\Sigma) - m_0(\Sigma)$
is equal to $m_1(\Xi) - m_0(\Xi)$ in this simple model.

Since we have assumed that the mixing angles for $\Sigma$ and
$\Lambda$ should be equal in the $\chi$QM$d$, the corresponding mixing
$\sin^2 \theta_\Sigma$ should be compared to the harmonic mean of
$\sin^2 \vartheta_\Sigma$ and $\sin^2 \vartheta_\Lambda$ in the toy
model, which is $\sin^2 \vartheta_{\Sigma\Lambda} \equiv 2 \sin^2
\vartheta_\Sigma \sin^2 \vartheta_\Lambda / (\sin^2 \vartheta_\Sigma +
\sin^2 \vartheta_\Lambda) \approx 0.27$.

Comparing the mixings in the toy model ($\sin^2 \vartheta_N \approx
0.19$, $\sin^2 \vartheta_{\Sigma\Lambda} \approx 0.27$, and $\sin^2
\vartheta_\Xi \approx 0.36$) with the ones obtained from the
$\chi$QM$d$ with $\alpha$ and $\beta$ free ($\sin^2 \theta_N \approx
0.11$, $\sin^2 \theta_\Sigma \approx 0.34$, and $\sin^2 \theta_\Xi
\approx 0.41$), we see that they are of the same order of magnitude
and they also appear in increasing order.

\section{Summary and Conclusions}
\label{sec:SC}

In this paper, we have studied the octet baryon magnetic moments in
the $\chi$QM with configuration mixing.  In particular, the
experimentally well established violation of the
Coleman--Glashow sum-rule cannot be reproduced in the $\chi$QM, no
matter how many SU(3) symmetry breaking parameters one introduces in
the Lagrangian~(\ref{eq:lag}).

As discussed, there are in principle two ways of overcoming this
problem, one is to let the quark magnetic moments vary between the
isomultiplets, and the other is to introduce symmetry breaking in the
wave functions of the octet baryons.  Taking the view, that the quarks
should have the same properties independently of in which baryon they
are, we are lead to choose the second alternative.

We considered two extensions of the $\chi$QM, one with quark-gluon
configuration mixing ($\chi$QM$g$), and one with quark-diquark
configuration mixing ($\chi$QM$d$).  The $\chi$QM$d$ with symmetry
breaking in the Lagrangian ($\alpha \approx 0.69$ and $\beta \approx
0.55$) led to $\Sigma_\mu \approx 0.52 \, \mu_N$, a value which lies
within the experimental errors.  The experimental value is $\Sigma_\mu
= (0.49 \pm 0.05) \, \mu_N$.  The values of the octet baryon magnetic
moments and the weak axial-vector form factor $g_A$ are also in very
good agreement with experiments.  The amount of quark-diquarks lies
between $11\%$ and $41\%$ in this model.

The introduction of a different set of symmetry breaking parameters in
the Lagrangian does not change the results significantly. The
violation of the Coleman--Glashow sum-rule is, in our models, solely
due to the configuration mixing parameters.

In conclusion, extensions of the $\chi$QM with configuration mixing of
quark-diquarks can explain the experimentally observed violation of
the Coleman--Glashow sum-rule for the octet baryon magnetic moments.

\acknowledgments

This work was supported by the Swedish Natural Science Research
Council (NFR), Contract No.  F-AA/FU03281-310.  Support for this work
was also provided by the Ernst Johnson Foundation (T.O.).

\appendix

\section{A Survey of the Chiral Quark Model}
\label{app:survey}

The Lagrangian ${\cal L}_I$ in Eq.~(\ref{eq:lag}), giving rise to GB
emission, will be specialized by putting $c_{\pi^0} = c_{\pi^+} =
c_{\pi^-} = 1$, $c_{K^+} = c_{K^-} = c_{K^0} = c_{\bar{K}^0} =
\alpha$, $c_\eta = \beta$, and $c_{\eta'} = \zeta$.  To find the quark
polarizations, we replace the GBs with their quark contents.  The
Lagrangian of the effective interaction is then given by
\begin{equation}
\hat{{\cal L}}_I = \sum_{q=u,d,s} \hat{{\cal L}}_q,
\end{equation}
where
\begin{displaymath}
	\hat{{\cal L}}_q = g_8 \bar{q} \sum_{q'=u,d,s}
	\hat{\Phi}_{qq'} q'.
\end{displaymath}
The transition of $q \rightarrow {\rm GB} + q' \rightarrow (q
\bar{q'})_0 + q'$, where $q' = u,d,s$, is described by the Lagrangian
$\hat{{\cal L}}_q$.  The matrix $\hat{\Phi}$ is
\begin{equation}
\hat{\Phi} = ( \hat{\Phi}_{qq'} ) = \left(
\begin{array}{ccc}
\phi_{uu} u \bar{u} + \phi_{ud} d \bar{d} + \phi_{us} s \bar{s} &
\varphi_{ud} u \bar{d} & \varphi_{us} u \bar{s}\\ \varphi_{du} d
\bar{u} & \phi_{du} u \bar{u} + \phi_{dd} d \bar{d} + \phi_{ds} s
\bar{s} & \varphi_{ds} d \bar{s}\\ \varphi_{su} s \bar{u} &
\varphi_{sd} s \bar{d} & \phi_{su} u \bar{u} + \phi_{sd} d \bar{d} +
\phi_{ss} s \bar{s}\\
\end{array}
\right), \label{Phihatt}
\end{equation}
where
$$
	\phi_{uu} = \phi_{dd} = \frac{1}{2} + \frac{\beta}{6} +
	\frac{\zeta}{3}, \hspace{3mm} \phi_{du} = \phi_{ud} = -
	\frac{1}{2} + \frac{\beta}{6} + \frac{\zeta}{3}, \hspace{3mm}
	\phi_{us} = \phi_{ds} = \phi_{su} = \phi_{sd} = -
	\frac{\beta}{3} + \frac{\zeta}{3},
$$
$$
	\phi_{ss} = \frac{2 \beta}{3} + \frac{\zeta}{3}, \hspace{3mm}
	\varphi_{ud} = \varphi_{du} = 1,
	\hspace{3mm}\mbox{and}\hspace{3mm} \varphi_{us} = \varphi_{ds}
	= \varphi_{su} = \varphi_{sd} = \alpha.
$$

The transition probability of $u$, $d$, and $s$ quarks can then be
expressed by the functions
\begin{eqnarray}
\vert \psi(u) \vert^2 &=& a \Bigg[ \left( 2 \phi_{uu}^2 + \phi_{ud}^2
+ \phi_{us}^2 + \varphi_{ud}^2 + \varphi_{us}^2 \right) \hat{u} +
\phi_{uu}^2 \hat{\bar{u}} \nonumber \\ &+& \left( \phi_{ud}^2 +
\varphi_{ud}^2 \right) \left( \hat{d} + \skew6\hat{\bar{d}} \right) +
\left( \phi_{us}^2 + \varphi_{us}^2 \right) \left( \hat{s} +
\skew2\hat{\bar{s}} \right) \Bigg],
\end{eqnarray}
\begin{eqnarray}
\vert \psi(d) \vert^2 &=& a \Bigg[ \left( \phi_{du}^2 + 2 \phi_{dd}^2
+ \phi_{ds}^2 + \varphi_{du}^2 + \varphi_{ds}^2 \right) \hat{d} +
\phi_{dd}^2 \skew6\hat{\bar{d}} \nonumber \\ &+& \left( \phi_{du}^2 +
\varphi_{du}^2 \right) \left( \hat{u} + \hat{\bar{u}} \right) + \left(
\phi_{ds}^2 + \varphi_{ds}^2 \right) \left( \hat{s} +
\skew2\hat{\bar{s}} \right) \Bigg],
\end{eqnarray}
and
\begin{eqnarray}
\vert \psi(s) \vert^2 &=& a \Bigg[ \left( \phi_{su}^2 + \phi_{sd}^2 +
2 \phi_{ss}^2 + \varphi_{su}^2 + \varphi_{sd}^2 \right) \hat{s} +
\phi_{ss}^2 \skew2\hat{\bar{s}} \nonumber \\ &+& \left( \phi_{su}^2 +
\varphi_{su}^2 \right) \left( \hat{u} + \hat{\bar{u}} \right) + \left(
\phi_{sd}^2 + \varphi_{sd}^2 \right) \left( \hat{d} +
\skew6\hat{\bar{d}} \right) \Bigg],
\end{eqnarray}
where $a \propto \vert g_8 \vert^2$ and the coefficients of the
$\hat{q}$ and $\skew3\hat{\bar{q}}$ should be interpreted as the
number of $q$ and $\bar{q}$ quarks, respectively.

The total probabilities of emission of a GB from $u$, $d$, and $s$
quarks are given by
\begin{equation}
\Sigma P_u = a \left( \phi_{uu}^2 + \phi_{ud}^2 + \phi_{us}^2 +
\varphi_{ud}^2 + \varphi_{us}^2 \right) = a \left( \frac{9 + \beta^2 +
2 \zeta^2}{6} + \alpha^2 \right),
\end{equation}
\begin{equation}
\Sigma P_d = a \left( \phi_{du}^2 + \phi_{dd}^2 + \phi_{ds}^2 +
\varphi_{du}^2 + \varphi_{ds}^2 \right) = a \left( \frac{9 + \beta^2 +
2 \zeta^2}{6} + \alpha^2 \right),
\end{equation}
and
\begin{equation}
\Sigma P_s = a \left( \phi_{su}^2 + \phi_{sd}^2 + \phi_{ss}^2 +
\varphi_{su}^2 + \varphi_{sd}^2 \right) = a \left( \frac{2 \beta^2 +
\zeta^2}{3} + 2 \alpha^2 \right).
\end{equation}
The total probability of no emission of GB from a $q$ quark is then
given by
\begin{equation}
P_q = 1 - \Sigma P_q.
\end{equation}

The antiquark numbers of the proton can be obtained from the
expression $2 P_u \hat{u} + P_d \hat{d} + 2 \vert \psi(u) \vert^2 +
\vert \psi(d) \vert^2$.  They are
\begin{equation}
\bar{u} = \frac{1}{12} \left[ \left( 2 \zeta + \beta + 1 \right)^2 +
20 \right] a,
\end{equation}
\begin{equation}
\bar{d} = \frac{1}{12} \left[ \left( 2 \zeta + \beta - 1 \right)^2 +
32 \right] a,
\end{equation}
and
\begin{equation}
\bar{s} = \frac{1}{3} \left[ \left( \zeta - \beta \right)^2 + 9
\alpha^2 \right] a.
\end{equation}

The spin structure of a baryon $B$ is described by the function
$\hat{B}$, which is defined by
\begin{equation}
\hat{B} \equiv \langle B^\uparrow \vert {\cal N} \vert B^\uparrow
\rangle,
\label{eq:spin_struct_def}
\end{equation}
where $\vert B^\uparrow \rangle$ is the wave function and ${\cal N}$
is the number operator
\begin{displaymath}
	{\cal N} = N_{u^\uparrow} \hat{u}^\uparrow + N_{u^\downarrow}
	\hat{u}^\downarrow + N_{d^\uparrow} \hat{d}^\uparrow +
	N_{d^\downarrow} \hat{d}^\downarrow + N_{s^\uparrow}
	\hat{s}^\uparrow + N_{s^\downarrow} \hat{s}^\downarrow.
\end{displaymath}

In the model with quark-gluon mixing ($\chi$QM$g$) the wave function
for $xxy$ baryons is
\begin{equation}
	\vert B^\uparrow (xxy) \rangle = \cos \theta_B \vert
	B_{1}^{\uparrow} (xxy) \rangle + \sin \theta_B \vert {\bf (}{B_{8}}
	(xxy)G{\bf )}^\uparrow \rangle.  \label{gluonwave}
\end{equation}
Simple calculations, using Eqs.~(\ref{eq:b1xxy}) and (\ref{eq:b8g}),
give
\begin{equation}
\langle B_{1}^{\uparrow} (xxy) \vert {\cal N} \vert B_{1}^{\uparrow}
(xxy) \rangle= \frac{5}{3} \hat{x}^\uparrow + \frac{1}{3}
\hat{x}^\downarrow + \frac{1}{3} \hat{y}^\uparrow + \frac{2}{3}
\hat{y}^\downarrow
\label{eq:b1xxy_2}
\end{equation}
and
\begin{equation}
\langle {\bf (}B_8 (xxy) G{\bf )}^\uparrow \vert {\cal N} \vert 
{\bf (}B_8 (xxy) G{\bf )}^\uparrow \rangle = \frac{8}{9}
\hat{x}^\uparrow + \frac{10}{9}
\hat{x}^\downarrow + \frac{4}{9} \hat{y}^\uparrow + \frac{5}{9}
\hat{y}^\downarrow.
\label{eq:b8Gxxy_2}
\end{equation}
The coefficients of the $\hat{q}^{\uparrow \downarrow}$ in the above
formulas should be interpreted as the number of $q^{\uparrow
\downarrow}$ quarks.

Using Eqs.~(\ref{eq:b1xxy_2}) and (\ref{eq:b8Gxxy_2}), and then making
the substitution
\begin{equation}
\hat{q}^\uparrow \rightarrow P_q \hat{q}^\uparrow + \vert \psi
(q^\uparrow) \vert^2,
\label{eq:subst_pil}
\end{equation}
for every quark, $q = u,d,s$, in the obtained formula, we get the spin
structure, after one interaction, as
\begin{eqnarray}
\hat{B}(xxy) &=& \cos^2 \theta_B \Bigg[ \frac{5}{3} \left( P_x
\hat{x}^\uparrow + \vert \psi(x^\uparrow) \vert^2 \right) +
\frac{1}{3} \left( P_x \hat{x}^\downarrow + \vert \psi(x^\downarrow)
\vert^2 \right) \nonumber \\ &+& \frac{1}{3} \left( P_y
\hat{y}^\uparrow + \vert \psi(y^\uparrow) \vert^2 \right) +
\frac{2}{3} \left( P_y \hat{y}^\downarrow + \vert \psi(y^\downarrow)
\vert^2 \right) \Bigg] \nonumber \\ &+& \sin^2 \theta_B \Bigg[
\frac{8}{9} \left( P_x \hat{x}^\uparrow + \vert \psi(x^\uparrow)
\vert^2 \right) + \frac{10}{9} \left( P_x \hat{x}^\downarrow + \vert
\psi(x^\downarrow) \vert^2 \right) \nonumber \\ &+& \frac{4}{9} \left(
P_y \hat{y}^\uparrow + \vert \psi(y^\uparrow) \vert^2 \right) +
\frac{5}{9} \left( P_y \hat{y}^\downarrow + \vert \psi(y^\downarrow)
\vert^2 \right) \Bigg],
\end{eqnarray}
where the functions $\vert \psi(q^{\uparrow \downarrow}) \vert^2$
describe the probability of emission of GBs, {\it i.e.} the
probability of transforming a $q^{\uparrow \downarrow}$ quark.

The probabilities of transforming $u$, $d$, and $s$ quarks with spin
up by one interaction can be expressed by the functions
\begin{eqnarray}
\vert \psi(u^\uparrow) \vert^2 &=& a \left[ \left( \phi_{uu}^2 +
\phi_{ud}^2 + \phi_{us}^2 \right) \hat{u}^\downarrow + \varphi_{ud}^2
\hat{d}^\downarrow + \varphi_{us}^2 \hat{s}^\downarrow \right]
\nonumber \\ &=& \frac{a}{6} \left(3+\beta^2+2 \zeta^2\right)
\hat{u}^\downarrow + a \hat{d}^\downarrow + a \alpha^2
\hat{s}^\downarrow, \label{eq:begin}
\end{eqnarray}
\begin{eqnarray}
\vert \psi(d^\uparrow) \vert^2 &=& a \left[ \left( \phi_{du}^2 +
\phi_{dd}^2 + \phi_{ds}^2 \right) \hat{d}^\downarrow + \varphi_{du}^2
\hat{u}^\downarrow + \varphi_{ds}^2 \hat{s}^\downarrow \right]
\nonumber \\ &=& a \hat{u}^\downarrow + \frac{a}{6} \left(3+\beta^2+2
\zeta^2\right) \hat{d}^\downarrow + a \alpha^2 \hat{s}^\downarrow,
\end{eqnarray}
and
\begin{eqnarray}
\vert \psi(s^\uparrow) \vert^2 &=& a \left[ \left( \phi_{su}^2 +
\phi_{sd}^2 + \phi_{ss}^2 \right) \hat{s}^\downarrow + \varphi_{su}^2
\hat{u}^\downarrow + \varphi_{sd}^2 \hat{d}^\downarrow \right]
\nonumber \\ &=& a \alpha^2 \hat{u}^\downarrow + a \alpha^2
\hat{d}^\downarrow + \frac{a}{3} \left(2\beta^2+\zeta^2\right)
\hat{s}^\downarrow.
\label{eq:end}
\end{eqnarray}
As before, the coefficient of $\hat{q}^\downarrow$ is the transition
probability to $q^\downarrow$.  We have here neglected the
quark-antiquark pair created by the GB, since it will not contribute
to the spin polarizations.

Similarly, in the model with quark-diquark mixing ($\chi$QM$d$), we
replace the wave function $|(B_{8}(xxy)G)^{\uparrow}\rangle$ by
$|B_{d}^{\uparrow}(xxy)\rangle$ in Eq.~(\ref{gluonwave}).  Using
Eq.~(\ref{eq:bd}), we find
\begin{equation}
\langle B_{d}^{\uparrow} (xxy) \vert {\cal N} \vert B_{d}^{\uparrow}
(xxy) \rangle = \hat{x}^\uparrow.
\end{equation}
After one interaction we then have
\begin{eqnarray}
\hat{B}(xxy) &=& \cos^2 \theta_B \Bigg[ \frac{5}{3} \left( P_x
\hat{x}^\uparrow + \vert \psi(x^\uparrow) \vert^2 \right) +
\frac{1}{3} \left( P_x \hat{x}^\downarrow + \vert \psi(x^\downarrow)
\vert^2 \right) \nonumber \\ &+& \frac{1}{3} \left( P_y
\hat{y}^\uparrow + \vert \psi(y^\uparrow) \vert^2 \right) +
\frac{2}{3} \left( P_y \hat{y}^\downarrow + \vert \psi(y^\downarrow)
\vert^2 \right) \Bigg] \nonumber \\ &+& \sin^2 \theta_B \left( P_x
\hat{x}^\uparrow + \vert \psi(x^\uparrow) \vert^2 \right).
\end{eqnarray}

The spin structure of the $\Lambda$ baryon after one interaction can
be obtained by a similar procedure like the one above for $xxy$
baryons.  The result is
\begin{eqnarray}
\hat{\Lambda}(uds) &=& \cos^2 \theta_\Sigma \Bigg[ P_u
\left(\frac{1}{2} \hat{u}^\uparrow + \frac{1}{2}
\hat{u}^\downarrow\right) + P_d \left(\frac{1}{2} \hat{d}^\uparrow +
\frac{1}{2} \hat{d}^\downarrow\right) + P_s \hat{s}^\uparrow \nonumber
\\ &+& \frac{1}{2} \vert \psi(u^\uparrow) \vert^2 + \frac{1}{2} \vert
\psi(u^\downarrow) \vert^2 + \frac{1}{2} \vert \psi(d^\uparrow)
\vert^2 + \frac{1}{2} \vert \psi(d^\downarrow) \vert^2 + \vert
\psi(s^\uparrow) \vert^2 \Bigg] \nonumber \\ &+& \sin^2 \theta_\Sigma
\Bigg[ P_u \left(\frac{1}{2} \hat{u}^\uparrow + \frac{1}{2}
\hat{u}^\downarrow\right) + P_d \left(\frac{1}{2} \hat{d}^\uparrow +
\frac{1}{2} \hat{d}^\downarrow\right) + P_s \left(\frac{1}{3}
\hat{s}^\uparrow + \frac{2}{3} \hat{s}^\downarrow\right) \nonumber \\
&+& \frac{1}{2} \vert \psi(u^\uparrow) \vert^2 + \frac{1}{2} \vert
\psi(u^\downarrow) \vert^2 + \frac{1}{2} \vert \psi(d^\uparrow)
\vert^2 + \frac{1}{2} \vert \psi(d^\downarrow) \vert^2 + \frac{1}{3}
\vert \psi(s^\uparrow) \vert^2 + \frac{2}{3} \vert \psi(s^\downarrow)
\vert^2 \Bigg]
\end{eqnarray}
in the $\chi$QM$g$ and
\begin{eqnarray}
\hat{\Lambda}(uds) &=& \cos^2 \theta_\Sigma \Bigg[ P_u
\left(\frac{1}{2} \hat{u}^\uparrow + \frac{1}{2}
\hat{u}^\downarrow\right) + P_d \left(\frac{1}{2} \hat{d}^\uparrow +
\frac{1}{2} \hat{d}^\downarrow\right) + P_s \hat{s}^\uparrow \nonumber
\\ &+& \frac{1}{2} \vert \psi(u^\uparrow) \vert^2 + \frac{1}{2} \vert
\psi(u^\downarrow) \vert^2 + \frac{1}{2} \vert \psi(d^\uparrow)
\vert^2 + \frac{1}{2} \vert \psi(d^\downarrow) \vert^2 + \vert
\psi(s^\uparrow) \vert^2 \Bigg] \nonumber \\ &+& \sin^2 \theta_\Sigma
\Bigg( \frac{1}{6} P_u \hat{u}^\uparrow + \frac{1}{6} P_d
\hat{d}^\uparrow + \frac{2}{3} P_s \hat{s}^\uparrow \nonumber \\ &+&
\frac{1}{6} \vert \psi(u^\uparrow) \vert^2 + \frac{1}{6} \vert
\psi(d^\uparrow) \vert^2 + \frac{2}{3} \vert \psi(s^\uparrow) \vert^2
\Bigg)
\end{eqnarray}
in the $\chi$QM$d$.

The spin polarization, $\Delta q^B$, where $q = u,d,s$, is defined as
\begin{equation}
\Delta q^B \equiv n_{q^\uparrow}(B) - n_{q^\downarrow}(B),
\end{equation}
where in the spin structure formulas $n_{q^\uparrow}(B)$ and
$n_{q^\downarrow}(B)$ are the coefficients of $\hat{q}^\uparrow$ and
$\hat{q}^\downarrow$, respectively, for the baryon $B$.  The spin
polarizations for the octet baryons are given in
Appendix~\ref{app:spin_pol}.

The magnetic moment of a baryon $B$ is determined from the expression
\begin{equation}
\mu(B) = \Delta u^B \mu_u + \Delta d^B \mu_d + \Delta s^B \mu_s.
\end{equation}

The total spin polarizations of the proton (the spin fraction carried
by the quarks in the proton) is given by
\begin{equation}
\Delta \Sigma = \Delta u^p + \Delta d^p + \Delta s^p.
\end{equation}

For the weak decay $n \to p + e^- + \bar{\nu}_e$ we can express the
weak axial-vector form factor, ${g_A}$, in terms of the spin
polarizations as
\begin{equation}
{g_A} = \Delta u^p - \Delta d^p.
\end{equation}

\section{Spin Polarizations}
\label{app:spin_pol}

\subsection{Spin polarizations in the $\chi$QM$g$}
\label{app:spin_pol_g}

The spin polarizations for the proton are given by
\begin{eqnarray}
\Delta u^p &=& \cos^2 \theta_N \left[ \frac{4}{3} - \frac{a}{3} \left(
7 + 4 \alpha^2 + \frac{4}{3} \beta^2 + \frac{8}{3} \zeta^2 \right)
\right] \nonumber \\ &+& \sin^2 \theta_N \left[ - \frac{2}{9} +
\frac{a}{9} \left( 5 + 2 \alpha^2 + \frac{2}{3} \beta^2 + \frac{4}{3}
\zeta^2 \right) \right] \\ \nonumber \\ \Delta d^p &=& \cos^2 \theta_N
\left[ - \frac{1}{3} - \frac{a}{3} \left( 2 - \alpha^2 - \frac{1}{3}
\beta^2 - \frac{2}{3} \zeta^2 \right) \right] \nonumber \\ &+& \sin^2
\theta_N \left[ - \frac{1}{9} + \frac{a}{9} \left( 4 + \alpha^2 +
\frac{1}{3} \beta^2 + \frac{2}{3} \zeta^2 \right) \right] \\ \nonumber
\\ \Delta s^p &=& \cos^2 \theta_N \left( - a \alpha^2 \right) + \sin^2
\theta_N \left( \frac{a}{3} \alpha^2 \right).
\end{eqnarray}

The spin polarizations for $\Sigma^+$ are given by
\begin{eqnarray}
\Delta u^{\Sigma^+} &=& \cos^2 \theta_\Sigma \left[ \frac{4}{3} -
\frac{a}{3} \left( 8 + 3 \alpha^2 + \frac{4}{3} \beta^2 + \frac{8}{3}
\zeta^2 \right) \right] \nonumber \\ &+& \sin^2 \theta_\Sigma \left[ -
\frac{2}{9} + \frac{a}{9} \left( 4 + 3 \alpha^2 + \frac{2}{3} \beta^2
+ \frac{4}{3} \zeta^2 \right) \right] \\ \nonumber \\ \Delta
d^{\Sigma^+} &=& \cos^2 \theta_\Sigma \left( \frac{a}{3} \left( -4 +
\alpha^2 \right) \right) + \sin^2 \theta_\Sigma \left( \frac{a}{9}
\left( 2 + \alpha^2 \right) \right) \\ \nonumber \\ \Delta
s^{\Sigma^+} &=& \cos^2 \theta_\Sigma \left[ - \frac{1}{3} -
\frac{a}{3} \left( 2 \alpha^2 - \frac{4}{3} \beta^2 - \frac{2}{3}
\zeta^2 \right) \right] \nonumber \\ &+& \sin^2 \theta_\Sigma \left[ -
\frac{1}{9} + \frac{a}{9} \left( 4 \alpha^2 + \frac{4}{3} \beta^2 +
\frac{2}{3} \zeta^2 \right) \right].
\end{eqnarray}

The spin polarizations for $\Xi^0$ are given by
\begin{eqnarray}
\Delta u^{\Xi^0} &=& \cos^2 \theta_\Xi \left[ - \frac{1}{3} +
\frac{a}{3} \left( 2 - 3 \alpha^2 + \frac{1}{3} \beta^2 + \frac{2}{3}
\zeta^2 \right) \right] \nonumber \\ &+& \sin^2 \theta_\Xi \left[ -
\frac{1}{9} + \frac{a}{9} \left( 2 + 3 \alpha^2 + \frac{1}{3} \beta^2
+ \frac{2}{3} \zeta^2 \right) \right] \\ \nonumber \\ \Delta d^{\Xi^0}
&=& \cos^2 \theta_\Xi \left( \frac{a}{3} \left( 1 - 4 \alpha^2 \right)
\right) + \sin^2 \theta_\Xi \left( \frac{a}{9} \left( 1 + 2 \alpha^2
\right) \right) \\ \nonumber \\ \Delta s^{\Xi^0} &=& \cos^2 \theta_\Xi
\left[ \frac{4}{3} - \frac{a}{3} \left( 7 \alpha^2 + \frac{16}{3}
\beta^2 + \frac{8}{3} \zeta^2 \right) \right] \nonumber \\ &+& \sin^2
\theta_\Xi \left[ - \frac{2}{9} + \frac{a}{9} \left( 5 \alpha^2 +
\frac{8}{3} \beta^2 + \frac{4}{3} \zeta^2 \right) \right].
\end{eqnarray}

The spin polarizations for $\Lambda$ are given by
\begin{eqnarray}
\Delta u^{\Lambda} &=& \cos^2 \theta_\Sigma \left( - a \alpha^2
\right) \nonumber \\ &+& \sin^2 \theta_\Sigma \left( \frac{a}{3}
\alpha^2 \right) \\ \nonumber \\ \Delta d^{\Lambda} &=& \cos^2
\theta_\Sigma \left( -a \alpha^2 \right) \nonumber \\ &+& \sin^2
\theta_\Sigma \left( \frac{a}{3} \alpha^2 \right) \\ \nonumber \\
\Delta s^{\Lambda} &=& \cos^2 \theta_\Sigma \left( 1 - \frac{a}{3}
\left( 6 \alpha^2 + 4 \beta^2 + 2 \zeta^2 \right) \right) \nonumber \\
&+& \sin^2 \theta_\Sigma \left[ - \frac{1}{3} + \frac{a}{3} \left( 2
\alpha^2 + \frac{4}{3} \beta^2 + \frac{2}{3} \zeta^2 \right) \right].
\end{eqnarray}

The spin polarizations for the other octet baryons are found from
isospin symmetry.

\subsection{Spin polarizations in the $\chi$QM$d$}
\label{app:spin_pol_d}

The spin polarizations for the proton are given by
\begin{eqnarray}
\Delta u^p &=& \cos^2 \theta_N \left[ \frac{4}{3} - \frac{a}{3} \left(
7 + 4 \alpha^2 + \frac{4}{3} \beta^2 + \frac{8}{3} \zeta^2 \right)
\right] \nonumber \\ &+& \sin^2 \theta_N \left[ 1 - a \left( 2 +
\alpha^2 + \frac{1}{3} \beta^2 + \frac{2}{3} \zeta^2 \right) \right]
\\ \nonumber \\ \Delta d^p &=& \cos^2 \theta_N \left[ - \frac{1}{3} -
\frac{a}{3} \left( 2 - \alpha^2 - \frac{1}{3} \beta^2 - \frac{2}{3}
\zeta^2 \right) \right] \nonumber \\ &+& \sin^2 \theta_N \left( - a
\right) \\ \nonumber \\ \Delta s^p &=& - a \alpha^2.
\end{eqnarray}

The spin polarizations for $\Sigma^+$ are given by
\begin{eqnarray}
\Delta u^{\Sigma^+} &=& \cos^2 \theta_\Sigma \left[ \frac{4}{3} -
\frac{a}{3} \left( 8 + 3 \alpha^2 + \frac{4}{3} \beta^2 + \frac{8}{3}
\zeta^2 \right) \right] \nonumber \\ &+& \sin^2 \theta_\Sigma \left[ 1
- a \left( 2 + \alpha^2 + \frac{1}{3} \beta^2 + \frac{2}{3} \zeta^2
\right) \right] \\ \nonumber \\ \Delta d^{\Sigma^+} &=& \cos^2
\theta_\Sigma \left( \frac{a}{3} \left( -4 + \alpha^2 \right) \right)
+ \sin^2 \theta_\Sigma \left( - a \right) \\ \nonumber \\ \Delta
s^{\Sigma^+} &=& \cos^2 \theta_\Sigma \left[ - \frac{1}{3} -
\frac{a}{3} \left( 2 \alpha^2 - \frac{4}{3} \beta^2 - \frac{2}{3}
\zeta^2 \right) \right] \nonumber \\ &+& \sin^2 \theta_\Sigma \left( -
a \alpha^2 \right).
\end{eqnarray}

The spin polarizations for $\Xi^0$ are given by
\begin{eqnarray}
\Delta u^{\Xi^0} &=& \cos^2 \theta_\Xi \left[ - \frac{1}{3} +
\frac{a}{3} \left( 2 - 3 \alpha^2 + \frac{1}{3} \beta^2 + \frac{2}{3}
\zeta^2 \right) \right] \nonumber \\ &+& \sin^2 \theta_\Xi \left( - a
\alpha^2 \right) \\ \nonumber \\ \Delta d^{\Xi^0} &=& \cos^2
\theta_\Xi \left( \frac{a}{3} \left( 1 - 4 \alpha^2 \right) \right) +
\sin^2 \theta_\Xi \left( - a \alpha^2 \right) \\ \nonumber \\ \Delta
s^{\Xi^0} &=& \cos^2 \theta_\Xi \left[ \frac{4}{3} - \frac{a}{3}
\left( 7 \alpha^2 + \frac{16}{3} \beta^2 + \frac{8}{3} \zeta^2 \right)
\right] \nonumber \\ &+& \sin^2 \theta_\Xi \left[ 1 - a \left( 2
\alpha^2 + \frac{4}{3} \beta^2 + \frac{2}{3} \zeta^2 \right) \right].
\end{eqnarray}

The spin polarizations for $\Lambda$ are given by
\begin{eqnarray}
\Delta u^{\Lambda} &=& \cos^2 \theta_\Sigma \left( - a \alpha^2
\right) \nonumber \\ &+& \sin^2 \theta_\Sigma \left[ \frac{1}{6} -
\frac{a}{6} \left( 3 + 5 \alpha^2 + \frac{\beta^2}{3} + \frac{2
\zeta^2}{3} \right) \right] \\ \nonumber \\ \Delta d^{\Lambda} &=&
\cos^2 \theta_\Sigma \left( - a \alpha^2 \right) \nonumber \\ &+&
\sin^2 \theta_\Sigma \left[ \frac{1}{6} - \frac{a}{6} \left( 3 + 5
\alpha^2 + \frac{\beta^2}{3} + \frac{2 \zeta^2}{3} \right) \right] \\
\nonumber \\ \Delta s^{\Lambda} &=& \cos^2 \theta_\Sigma \left( 1 -
\frac{a}{3} \left( 6 \alpha^2 + 4 \beta^2 + 2 \zeta^2 \right) \right)
\nonumber \\ &+& \sin^2 \theta_\Sigma \left[ \frac{2}{3} - \frac{a}{3}
\left( 5 \alpha^2 + \frac{8 \beta^2}{3} + \frac{4 \zeta^2}{3} \right)
\right].
\end{eqnarray}

The spin polarizations for the other octet baryons are found from
isospin symmetry.

\onecolumn \widetext

\begin{table}
\caption{Parameter values obtained in the different fits.  The
subscript $_{\alpha \beta}$ in a model name indicates that the
parameters $\alpha$ and $\beta$ were allowed to vary in the fit.
Hyphen (-) indicates that the parameter was not defined in the fit.
$(1)$ means that the parameter was not free in the fit, but put to
$1$.  The magnetic moment of the $d$ quark, $\mu_d$, is given in units
of the nuclear magneton, $\mu_N$.}
\begin{tabular}{lrrrrrrrrr}
Parameter & NQM & NQM$g$ & NQM$d$ & $\chi$QM & $\chi_{\alpha \beta}$QM
& $\chi$QM$g$ & $\chi_{\alpha \beta}$QM$g$ & $\chi$QM$d$ &
$\chi_{\alpha \beta}$QM$d$\\ \hline $\mu_d$ & $-0.91$ & $-1.15$ &
$-1.09$ & $-1.35$ & $-1.23$ & $-1.40$ & $-1.24$ & $-1.42$ & $-1.27$\\
$\alpha$ & - & - & - & $(1)$ & $0.52$ & $(1)$ & $0.70$ & $(1)$ &
$0.69$\\ $\beta$ & - & - & - & $(1)$ & $0.99$ & $(1)$ & $0.73$ & $(1)$
& $0.55$\\ $\sin^2 \theta_N$ & - & $0.18$ & $0.39$ & - & - & $0.00$ &
$0.00$ & $0.00$ & $0.11$\\ $\sin^2 \theta_\Sigma$ & - & $0.20$ &
$0.65$ & - & - & $0.05$ & $0.04$ & $0.25$ & $0.34$\\ $\sin^2
\theta_\Xi$ & - & $0.24$ & $0.46$ & - & - & $0.11$ & $0.11$ & $0.33$ &
$0.41$\\
\end{tabular}
\label{tab:fit_data}
\end{table}

\begin{table}
\caption{Octet baryon magnetic moments, $g_A$, and $\Sigma_\mu$.  The
subscript $_{\alpha \beta}$ in a model name indicates that the
parameters $\alpha$ and $\beta$ were allowed to vary in the fit.  The
octet baryon magnetic moments and $\Sigma_\mu$ are given in units of
the nuclear magneton, $\mu_N$.  The experimental values have been
obtained from Ref.~\protect\cite{barn96}.}
\begin{tabular}{lcrrrrrrrrr}
Quantity & Expt.  values & NQM & NQM$g$ & NQM$d$ & $\chi$QM &
$\chi_{\alpha \beta}$QM & $\chi$QM$g$ & $\chi_{\alpha \beta}$QM$g$ &
$\chi$QM$d$ & $\chi_{\alpha \beta}$QM$d$\\ \hline $\chi^2$ & & $0.28$
& $0.14$ & $0.081$ & $0.12$ & $0.075$ & $0.082$ & $0.055$ & $0.031$ &
$0.0032$\\ \hline $\mu(p)$ & $2.79 \pm 0.00$ & $2.72$ & $2.77$ &
$2.85$ & $2.67$ & $2.65$ & $2.76$ & $2.74$ & $2.80$ & $2.76$\\
$\mu(n)$ & $-1.91 \pm 0.00$ & $-1.81$ & $-1.89$ & $-1.76$ & $-1.86$ &
$-1.94$ & $-1.92$ & $-1.96$ & $-1.95$ & $-1.95$\\ $\mu(\Sigma^+)$ &
$2.46 \pm 0.01$ & $2.61$ & $2.56$ & $2.53$ & $2.57$ & $2.52$ & $2.52$
& $2.49$ & $2.48$ & $2.46$\\ $\mu(\Sigma^-)$ & $-1.16 \pm 0.03$ &
$-1.01$ & $-0.95$ & $-1.14$ & $-1.05$ & $-1.15$ & $-1.02$ & $-1.07$ &
$-1.07$ & $-1.15$\\ $\mu(\Xi^0)$ & $-1.25 \pm 0.01$ & $-1.41$ &
$-1.38$ & $-1.25$ & $-1.45$ & $-1.41$ & $-1.35$ & $-1.35$ & $-1.24$ &
$-1.25$\\ $\mu(\Xi^-)$ & $-0.65 \pm 0.00$ & $-0.50$ & $-0.41$ &
$-0.66$ & $-0.55$ & $-0.49$ & $-0.48$ & $-0.48$ & $-0.61$ & $-0.67$\\
$\mu(\Lambda)$ & $-0.61 \pm 0.00$ & $-0.60$ & $-0.56$ & $-0.45$ &
$-0.65$ & $-0.62$ & $-0.63$ & $-0.64$ & $-0.59$ & $-0.61$\\ $g_A$ &
$1.26 \pm 0.00$ & $\frac{5}{3}$ & $1.35$ & $1.41$ & $1.12$ & $1.24$ &
$1.12$ & $1.26$ & $1.12$ & $1.24$\\[1mm] $\Sigma_\mu$ & $0.49 \pm
0.05$ & $0$ & $0.17$ & $0.36$ & $0$ & $0$ & $0.28$ & $0.27$ & $0.55$ &
$0.52$\\
\end{tabular}
\label{tab:octet}
\end{table}

\begin{table}
\caption{Spin polarizations for the proton.  The subscript $_{\alpha
\beta}$ in a model name indicates that the parameters $\alpha$ and
$\beta$ were free in the fit.}
\begin{tabular}{lrrrrrrrrr}
Quantity & NQM & NQM$g$ & NQM$d$ & $\chi$QM & $\chi_{\alpha \beta}$QM
& $\chi$QM$g$ & $\chi_{\alpha \beta}$QM$g$ & $\chi$QM$d$ &
$\chi_{\alpha \beta}$QM$d$\\ \hline $\Delta u^p$ & $\frac{4}{3}$ &
$1.05$ & $1.20$ & $0.79$ & $0.89$ & $0.79$ & $0.91$ & $0.79$ &
$0.91$\\ $\Delta d^p$ & $-\frac{1}{3}$ & $-0.29$ & $-0.20$ & $-0.32$ &
$-0.35$ & $-0.32$ & $-0.35$ & $-0.32$ & $-0.33$\\ $\Delta s^p$ & $0$ &
$0$ & $0$ & $-0.10$ & $-0.03$ & $-0.10$ & $-0.05$ & $-0.10$ &
$-0.05$\\ $\Delta \Sigma$ & $1$ & $0.76$ & $1$ & $0.37$ & $0.52$ &
$0.37$ & $0.51$ & $0.37$ & $0.53$\\
\end{tabular}
\label{tab:data}
\end{table}

\end{document}